\documentclass[twocolumn,showpacs,superscriptaddress,prb]{revtex4-1}
\usepackage{amssymb}
\usepackage{amsmath}
\usepackage[dvips]{graphicx}
\usepackage{dcolumn}
\usepackage{bm}
\usepackage{color}

\begin{document}

\title{Electric field-driven topological phase switching and skyrmion lattice metastability in magnetoelectric Cu$_{2}$OSeO$_{3}$}

\author{J.\,S.\,White}\email{jonathan.white@psi.ch}
\affiliation{Laboratory for Neutron Scattering and Imaging (LNS), Paul Scherrer Institut (PSI), CH-5232 Villigen, Switzerland}

\author{I.\,\v{Z}ivkovi\'{c}}
\affiliation{Laboratory for Quantum Magnetism (LQM), \'{E}cole Polytechnique F\'{e}d\'{e}rale de Lausanne (EPFL), CH-1015 Lausanne, Switzerland}

\author{A.\,J.\,Kruchkov}
\affiliation{Department of Physics, Harvard University, Cambridge MA 02138, USA}

\author{M.\,Bartkowiak}
\affiliation{Laboratory for Scientific Developments and Novel Materials (LDM), Paul Scherrer Institut (PSI), CH-5232 Villigen, Switzerland}

\author{A.\,Magrez}
\affiliation{Crystal Growth Facility, \'{E}cole Polytechnique F\'{e}d\'{e}rale de Lausanne (EPFL), CH-1015 Lausanne, Switzerland}


\author{H.\,M.\,R\o nnow}
\affiliation{Laboratory for Quantum Magnetism (LQM), \'{E}cole Polytechnique F\'{e}d\'{e}rale de Lausanne (EPFL), CH-1015 Lausanne, Switzerland}

\date{\today}

\begin{abstract}
Due to their topological protection and nanometric size, magnetic skyrmions are anticipated to form components of new high-density memory technologies. In metallic systems skyrmion manipulation is achieved easily under a low density electric current flow, although the inevitable thermal dissipation ultimately limits the energy efficacy of potential applications. On the other hand, a near dissipation-free skyrmion and skyrmion phase manipulation is expected by using electric \emph{fields}, thus meeting better the demands of an energy-conscious society. In this work on an insulating chiral magnet Cu$_{2}$OSeO$_{3}$ with magnetoelectric coupling, we use neutron scattering to demonstrate directly i) the creation of metastable skyrmion states over an extended range in magnetic field and temperature ($T$), and ii) the in-situ electric field-driven switching between topologically distinct phases; the skyrmion phase and a competing non-topological cone phase. For our accessible electric field range, the phase switching is achieved in a high temperature regime, and the remnant ($E=$~0) metastable skyrmion state is thermally volatile with an exponential lifetime on hour timescales. Nevertheless, by taking advantage of the demonstrably longer-lived metastable skyrmion states at lower temperatures, a truly non-volatile and near dissipation-free topological phase change memory function is promised in magnetoelectric chiral magnets.
\end{abstract}

\maketitle
\section{Introduction}
Magnetic skyrmions are nanoscale spin vortex-like objects which hold clear promise for use in spintronics applications by virtue of their topological protection.~\citep{Nag13,Wie16,Kan17} Recently, intense activity has been devoted to the creation and exploitation of metastable skyrmion states in non-centrosymmetric magnets.~\cite{Mun10,Mil13,Bau16,Wil17,Oik16,Mak17,Nak17,Kag17a,Kag17,Kar16,Mor17,Kar17,Ban17,Oka17} Such states are created by thermally quenching a high temperature ($T$) equilibrium skyrmion phase to lower $T$, and avoiding kinetically the ubiquitous first-order phase transition to adjacent topologically trivial equilibrium phases at lower $T$.~\cite{Nag13} The cooling rate required to create metastable skyrmion states varies strongly with system homogeneity. In stoichiometric MnSi rapid cooling rates of $\sim10^{2}$~K/s are needed to create metastable skyrmion states,~\cite{Oik16,Nak17,Kag17a} while in MnSi under inhomogeneous strain,~\cite{Rit13} or alloys with structural disorder such as Fe$_{x}$Co$_{1-x}$Si~\cite{Mun10,Mil13,Bau16,Wil17} and (Co$_{0.5}$Zn$_{0.5}$)$_{20-x}$Mn$_{x}$ ($x$=2,4),~\cite{Kar16,Mor17,Kar17} the cooling rates of $\sim$10$^{-3}$-10$^{1}$~K/s provided by standard laboratory equipment are sufficient. The interest in metastable skyrmion states stems from their potential to display properties not shown by equilibrium skyrmion phases, and provide a platform for exploring physical mechanisms that are directly relevant for skyrmion-based applications, including topological phase stability, decay, creation and annihilation.\par

To explore these important issues, studies of insulating skyrmion host materials are particularly attractive. The non-centrosymmetry of the crystal lattice endows these compounds with a magnetoelectric (ME) coupling which, in turn mediates a direct, energy-efficient electric ($E$) field control of the skyrmion spin texture. However, the number of bulk insulating skyrmion hosts remains extremely limited. The known candidates include recently discovered polar magnets that host N\'{e}el-type skyrmions, namely GaV$_{4}$(S,Se)$_{8}$ ($T_{c}$$\leq$18~K)~\citep{Kez15,Whi18,Fuj17,Bor17} and VOSe$_{2}$O$_{5}$ ($T_{c}$=7.5~K),~\citep{Kur17} and the unique Bloch-type skyrmion host Cu$_{2}$OSeO$_{3}$ ($T_{c}$=58~K) with chiral cubic $P$2$_{1}$3 structure~\citep{Sek12,Ada12}. Of the polar magnets, ME coupling below $T_{c}$ has been reported only for GaV$_{4}$(S,Se)$_{8}$,~\citep{Ruf15b,Ruf17} with the role of this ME coupling on skyrmion phase stability and metastability an open question.\par

In the present work we focus our attention on the chiral magnet Cu$_{2}$OSeO$_{3}$ which has a dielectric constant $\epsilon_{r}$$\sim$6~\citep{Ziv12} and a well-established ME coupling below $T_{c}$.~\cite{Bos08,Bel10,Bel12,Mai12,Sek12,Sek12b,Whi12,Ziv12,Omr14,Whi14,Ruf15,Mil16} Although Cu$_{2}$OSeO$_{3}$ displays magnet ordering well below room temperature, it is likely that the room temperature skyrmion-hosting insulators anticipated to be discovered in the near-future will have a similar chiral cubic non-centrosymmetric lattice symmetry. Such materials are expected to host Bloch-type skyrmions and display a qualitatively similar magnetoelectric coupling effect as Cu$_{2}$OSeO$_{3}$. Therefore insights obtained from studying Cu$_{2}$OSeO$_{3}$ can be anticipated to be of direct relevance to future applications at room temperature. In this context, a detailed understanding of the $E$ field response of both equilibrium and metastable Bloch-type skyrmion states in the only available chiral cubic insulating system Cu$_{2}$OSeO$_{3}$ is of clear importance for the nascent research field of `skyrmionics'. Moreover, recent theoretical expectation for the $E$ field-induced creation and annihilation of skyrmions in insulators,~\citep{Wat16,Moc16} and also the $E$ field-control of skyrmion phase stability,~\citep{Kru17} motivate the present work which focuses on functionalities that are attractive for applications.\par


In fact, in Cu$_{2}$OSeO$_{3}$ metastable skyrmion states were recently reported to be created in both bulk~\cite{Oka16} and strained thin plate samples.~\cite{Oka17} In the bulk sample, the metastable skyrmion state was created by magnetoelectric field-cooling (MEFC); that is, quenching the equilibrium skyrmion phase to low $T$ in simultaneously applied $E$ and magnetic fields, under cooling rates of $\sim$10$^{0}$-10$^{1}$~K/s.~\cite{Oka16} In addition, the extent of the equilibrium skyrmion phase space can be either enhanced or suppressed dependent on the polarity of an applied $E$ field.~\cite{Oka16,Kru17} Taking together these observations, parametric conditions can be identified whereby Cu$_{2}$OSeO$_{3}$ presents a platform for an $E$ field-driven topological phase switching between bistable skyrmion and topologically trivial helical/conical phases. Suggestive evidence for the hysteretic switching between these two phases was presented by Okamura~\emph{et al},~\cite{Oka16} though only indirectly by measurement of bulk susceptibility. From this earlier work, the $E$ field-driven exchange of competing phase volume fractions cannot be evidenced directly, and the temporal stability (volatility) of the remnant metastable skyrmion state in zero bias $E$ field was not characterized.\par

Here we apply small-angle neutron scattering (SANS) as a powerful, microscopic probe of metastable skyrmion states and topological phase switching in bulk Cu$_{2}$OSeO$_{3}$. By using a standard MEFC procedure, metastable skyrmion states are created that exist over an extended parameter space compared with the typically narrow extent of the equilibrium skyrmion phase, including base $T$, and an extended range of applied magnetic field ($\mu_{0}H$) - see Fig.~\ref{fig:phase_diagrams}. At suitable $\mu_{0}H$ and $T$ points just outside the equilibrium skyrmion phase at $E$=0, the \emph{in-situ} $E$ field-driven phase switching between topologically distinct skyrmion and conical phases is achieved, though the metastable skyrmion state at remnance ($E$=0) is thermally volatile with an exponential lifetime on the hour timescale. Nevertheless, by creating longer-lived remnant metastable states at lower $T$, we show chiral magnets to promise potentially useful phase change memory functionalities based on ME coupling that should in principle exist also at room temperature.\par

\section{Experimental}
Single crystals of Cu$_{2}$OSeO$_{3}$ ($T_{c}$=58~K) were prepared by chemical vapor transport.~\citep{Bel10} A crystal of volume $3.0\times2.0\times0.50$~mm$^{3}$ and mass 6~mg was mounted onto a bespoke $E$ field sample stick.~\citep{Bar14} A dc $E$ field was applied along the thin axis of the sample $\parallel$$[111]$. The horizontal plane was defined by orthogonal $[111]-[1\bar{1}0]$ cubic axes, with $[\bar{1}\bar{1}2]$ vertical. At cryogenic $T$s, the experimental $E$ field ranged from -2.5~V/$\mu$m$<$$E$$<$+5.0~V/$\mu$m, with signs of arcing detected outside this range. The $E$-field stick was installed into a horizontal field cryomagnet and a field geometry of $E$$\parallel$$\mu_{0}H$$\parallel$$[111]$ was used throughout. In this field geometry the directions of $E$, and the $\mu_{0}H$-induced electric polarization coincide,~\citep{Sek12} providing the optimal conditions for the direct $E$ field control of the equilibrium skyrmion phase stability.~\citep{Kru17}\par

SANS measurements were done using the SANS-II beamline, at Paul Scherrer Institut, Switzerland.~\citep{Mis1} Two experimental geometries were employed; firstly $\mu_{0}H$$\parallel$$\textbf{k}_{i}$, where $\textbf{k}_{i}$ is the incident neutron wavevector, and secondly $\mu_{0}H$$\perp$$\textbf{k}_{i}$. In the first geometry the skyrmion phase can be studied, since the skyrmions form a quasi-long range ordered hexagonal skyrmion lattice (SkL) that is always observed as a six-fold pattern of diffraction peaks all with $\textbf{q}$$\perp$$\mu_{0}H$. In this geometry no SANS signal can be detected due to the helical ($\textbf{q}_{h}$$\parallel$$\langle 100\rangle$) or conical phases ($\textbf{q}_{c}$$\parallel$$\mu_{0}H$), since the associated $q$-vectors lie well out of the SANS detector plane. Instead, SANS signal from these phases can be observed in the second geometry. SANS data collection time at each stabilized $T$, $\mu_{0}H$ and $E$ varied from 10-20~mins.

\section{Results and Discussion}
\subsection{Equilibrium magnetic phases and metastable skyrmion state}

\begin{figure}[t]
\includegraphics[width=0.48\textwidth]{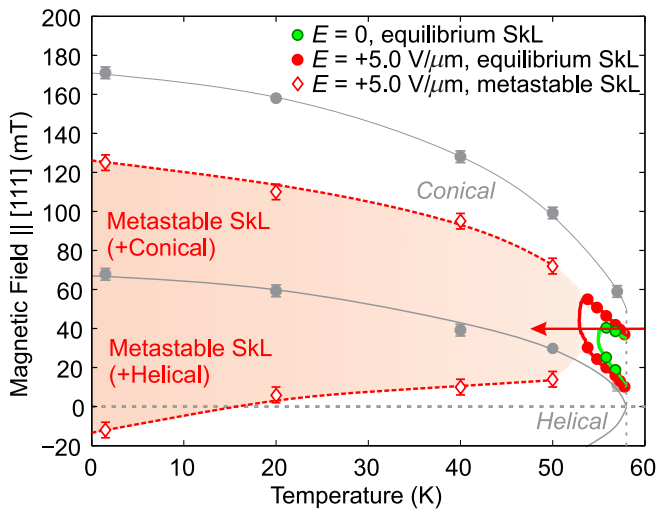}
\caption{The phase diagram of bulk Cu$_{2}$OSeO$_{3}$ for $\mu_{0}H$$\parallel$$[111]$ inferred from SANS. For $E$=0, the equilibrium phase boundaries are denoted by gray (green) symbols and lines for the helical and conical (skyrmion) phases. The extent of the equilibrium skyrmion phase under $E$=+5.0~V/$\mu$m is shown by filled red circles.~\cite{Kru17} Empty diamonds and dashed lines denote the extent of the metastable skyrmion state created by MEFC through $T_{c}$ in both $E$=+5.0~V/$\mu$m and $\mu_{0}H=$~40~mT (route shown by the red arrow). All lines are guides for the eye.}
\label{fig:phase_diagrams}
\end{figure}

Fig.~\ref{fig:phase_diagrams} shows the equilibrium phase diagram with $\mu_{0}H$$\parallel$$[111]$ and $E$=0 determined by SANS. In our experimental geometry with $\mu_{0}H$ applied normal to surface of the plate-like sample, the resulting demagnetizing field elevates the overall magnetic field scale of the phase diagram with respect to other phase diagrams reported elsewhere for bulk Cu$_{2}$OSeO$_{3}$.~\citep{Sek12c,Whi14,Omr14,Lev16} Nevertheless, the general structure of our phase diagram is identical to those reported earlier, and furthermore the same as those of other bulk Dzyaloshinkii-Moriya magnets with chiral cubic $P$2$_{1}$3~\cite{Muh09,Wil11,Mun10} or $P$4$_{1}$32/$P$4$_{3}$32 symmetries.~\cite{Tok15,Kar16,Kar17} A multi-domain single-$\textbf{q}$ helical phase is stable for internal fields ($B$) $0<B<B_{c1}(T)$, and a single-$\textbf{q}$ domain spin-flop or conical phase is stable for $B_{c1}(T)<B<B_{c2}(T)$. The topological skyrmion phase occupies a narrow interval in finite $\mu_{0}H$ and $T$ just below $T_{c}$, and is bordered at low $T$ by the non-topological conical phase. The helical and conical phase boundaries were determined from $\mu_{0}H$-increasing scans of the associated SANS integrated intensities done in the $\mu_{0}H$$\perp$$\textbf{k}_{i}$ geometry after zero field-cooling (ZFC). From data obtained on the same crystal reported in Ref.~\onlinecite{Kru17}, the extents of the equilibrium skyrmion phases for $E=$~0 and under $E=$+5.0~V/$\mu$m were determined from $\mu_{0}H$-increasing scans done in the $\mu_{0}H$$\parallel$$\textbf{k}_{i}$ geometry.\par





\begin{figure}[t]
\includegraphics[width=0.48\textwidth]{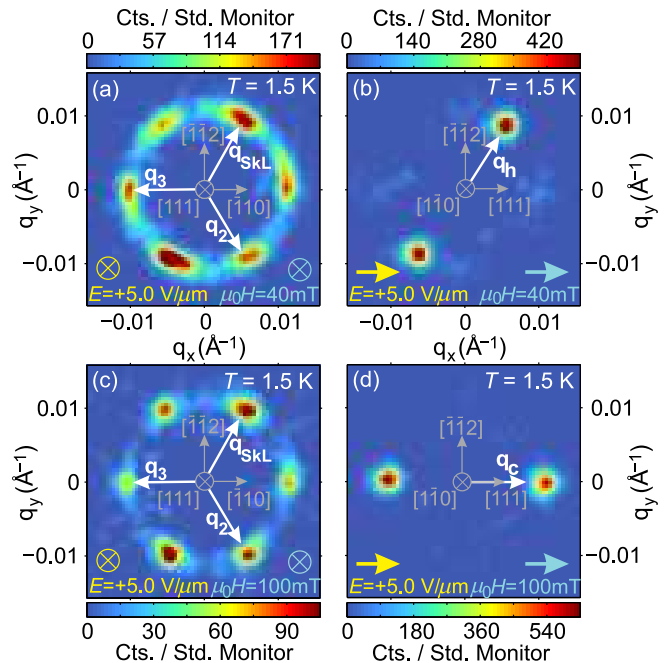}
\caption{SANS patterns obtained in (a) the $\mu_{0}H$$\parallel$$\textbf{k}_{i}$ and (b) the $\mu_{0}H$$\perp$$\textbf{k}_{i}$ configurations at $T=$~1.5~K, after the MEFC. (c) and (d) show corresponding SANS patterns collected after a subsequent magnetic field ramp to $\mu_{0}H=$~100~mT. The geometry of applied fields was always maintained to be $E$$\parallel$$\mu_{0}H$$\parallel$$[111]$. In (a) and (c) the propagation vectors denoted $\textbf{q}_{\rm SkL}$, $\textbf{q}_{\rm 2}$ and $\textbf{q}_{\rm 3}$ all describe scattering from the SkL. Note that both $\pm$\textbf{q} each give a Bragg spot. In (b) $\textbf{q}_{\rm h}$ denotes scattering from the helical phase, while in (d) $\textbf{q}_{\rm c}$ denotes scattering from the conical phase. All images correspond to views of the detector as seen from the sample.}
\label{fig:SANS_patterns}
\end{figure}

The metastable skyrmion state was created in the sample by MEFC from 70~K to the target $T$ using a cooling rate of -0.02~K/s, and in simultaneously applied fields of $E=$+5.0~V/$\mu$m and $\mu_{0}H=$~40~mT - the route shown by the red arrow in Fig.~\ref{fig:phase_diagrams}. Fig.~\ref{fig:SANS_patterns}(a) shows the SANS pattern obtained from the resulting state at $T$=1.5~K in the $\mu_{0}H$$\parallel$$\textbf{k}_{i}$ geometry. The six-fold pattern confirms the existence of a metastable SkL. At the same time in the $\mu_{0}H$$\perp$$\textbf{k}_{i}$ geometry [Fig.~\ref{fig:SANS_patterns}(b)] a twofold SANS pattern is observed due to a domain of the equilibrium helical phase with $\textbf{q}_{h}$$\parallel$$[001]$ that co-exists with the metastable SkL. While maintaining $E$, increasing $\mu_{0}H$ to 100~mT shows the metastable SkL spot intensities to fall [Fig.~\ref{fig:SANS_patterns}(c)], while in the $\mu_{0}H$$\perp$$\textbf{k}_{i}$ geometry the coexisting phase has undergone a magnetic transition to the conical phase with $\textbf{q}_{c}$$\parallel$$\mu_{0}H$ [Fig.~\ref{fig:SANS_patterns}(d)].\par

\begin{figure}[t]
\includegraphics[width=0.48\textwidth]{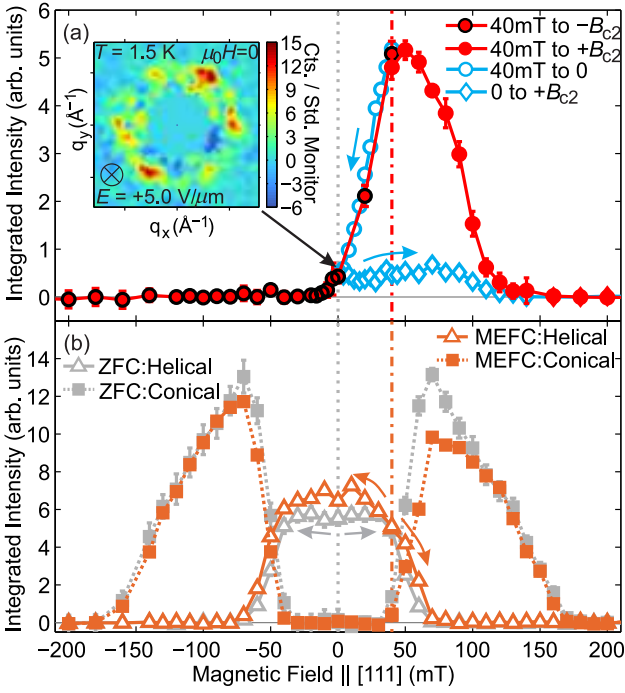}
\caption{(a) $\mu_{0}H$-scans of the SANS integrated intensity from the metastable SkL at $T$=1.5~K. Red symbols denote scans each carried out after a separate MEFC. Blue symbols denote scans done successively after a single MEFC as indicated by the blue arrows. The inset shows the SANS pattern obtained from the metastable SkL at $\mu_{0}H$=0. (b) Corresponding $\mu_{0}H$-scans of the SANS intensity due to the helical and conical phases observed in $\mu_{0}H\perp\textbf{k}_{i}$ geometry. Orange symbols denote the intensities of the two phases after MEFC. The scans up or down in $\mu_{0}H$ all begin at 40~mT. Gray symbols denote data obtained after ZFC and so the $\mu_{0}H$ scans begin at 0~mT. From similar scans done after ZFC but at higher $T$s (data not shown), the overall extent of the equilibrium phase diagram shown in Fig.~\ref{fig:phase_diagrams} was determined. The data shown in (a) and (b) were all obtained in the fixed $E$$\parallel$$\mu_{0}H$$\parallel$$[111]$ field geometry.}
\label{fig:ScANS_metastable}
\end{figure}

\begin{figure*}[t]
\includegraphics[width=0.95\textwidth]{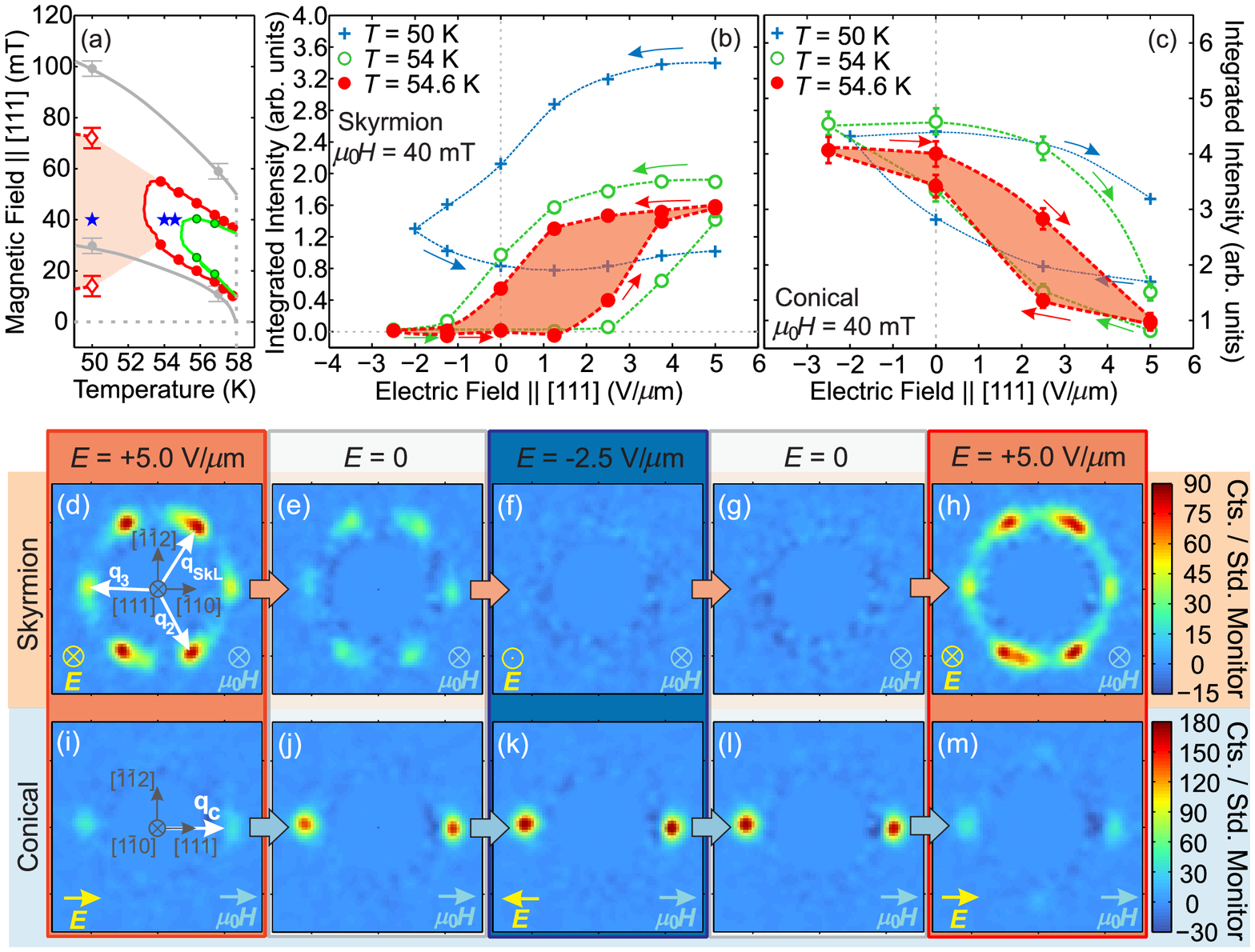}
\caption{(a) The high $T$ portion of the phase diagram shown in Fig.~1. Blue stars denote ($T$,$\mu_{0}H$) locations where $E$ field sweeping measurements were performed after an initial MEFC. In the fixed $E$$\parallel$$\mu_{0}H$$\parallel$$[111]$ field geometry, (b) and (c) show the $E$ field sweeping dependence of the SANS integrated intensities for (b) the SkL in the $H$$\parallel$$\textbf{k}_{i}$ geometry and, (c) the competing conical state in the $H$$\perp$$\textbf{k}_{i}$ geometry. The dashed lines in (b) and (c) are guides for the eye. Sequences of SANS patterns obtained during the $E$ field sweeping at fixed $T=$54.6~K, and $\mu_{0}H$=40~mT are shown in (d)-(h) for the SkL and (i)-(m) conical phases. Each row of SANS images maintains a common intensity scale.}
\label{fig:Efield_sweeping}
\end{figure*}

Fig.~\ref{fig:ScANS_metastable}(a) shows $\mu_{0}H$-increasing and decreasing scans of the metastable SkL SANS integrated intensity at $T$=1.5~K, and always under $E=$+5.0~V/$\mu$m. The metastable SkL survives over an extended field range $-12(5)$~mT$<\mu_{0}H<+125(5)$~mT, including $\mu_{0}H$=0 as shown in the inset of Fig.~\ref{fig:ScANS_metastable}(a). Further scans were carried out to explore the metastable SkL in the vicinity of $\mu_{0}H$=0 [blue data symbols in Fig.~\ref{fig:ScANS_metastable}(a)]. The intensity lost upon ramping down $\mu_{0}H$ from 40~mT to 0 is not recovered when subsequently increasing $\mu_{0}H$ again. These data show that once metastable skyrmions are destroyed, they cannot be renucleated without raising the $T$ towards the range of the equilibrium skyrmion phase near $T_{c}$. Similar $\mu_{0}H$-increasing and decreasing scans done after MEFC to higher $T$s (data not shown) allowed us to determine the parametric extent of the metastable skyrmion state shown in Fig.~\ref{fig:phase_diagrams}.\par

In Fig.~\ref{fig:ScANS_metastable}(b) we show $\mu_{0}H$-scans of the SANS integrated intensities from the coexisting helical and conical phases at 1.5~K after MEFC. A clear asymmetry between $\mu_{0}H$-increasing and decreasing sweeps is observed for the conical phase after MEFC. This is unlike after ZFC with $E$=0, where the $\mu_{0}H$-increasing and decreasing sweeps are symmetric around $\mu_{0}H$=0 as expected. We find that for the $\mu_{0}H$-decreasing case, all MEFC and ZFC curves are very similar, particularly when $\mu_{0}H<0$ and there is no co-existing metastable SkL. In contrast, for the $\mu_{0}H$-increasing cases the conical phase intensity after MEFC is less than for the ZFC case over the $\mu_{0}H$ range where it coexists with the metastable SkL. By taking into account this difference in conical phase intensity between the MEFC and ZFC scans, the metastable SkL volume fraction created by MEFC is estimated to be 25(5)\%.

Inspecting the SANS patterns behind the $\mu_{0}H$-scan data shown in Fig.~\ref{fig:ScANS_metastable}(a) shows the metastable SkL remains always hexagonally coordinated. This contrasts with the cases of Co$_{8}$Zn$_{8}$Mn$_{4}$~\cite{Kar16,Mor17} and MnSi~\cite{Nak17} where transformations to skyrmion states with an average fourfold symmetry take place as $\mu_{0}H$ is decreased. In those cases, the $\mu_{0}H$ was applied along a cubic axis with either fourfold (Co$_{8}$Zn$_{8}$Mn$_{4}$) or twofold (MnSi) rotation symmetry. In these situations, symmetry may allow the stabilization of a SkL coordination with a respective four or twofold rotational symmetry axis coincident with $\mu_{0}H$. Here, for Cu$_{2}$OSeO$_{3}$ with $\mu_{0}H$ aligned along a threefold $[111]$ axis, SkL coordinations with a two or fourfold rotation axis coincident with $\mu_{0}H$ are less favoured by symmetry, thus explaining the persistence of the hexagonal coordination for all $\mu_{0}H$. Studies with $\mu_{0}H$ aligned with a cubic axis can clarify if a metastable SkL transformation may yet exist in Cu$_{2}$OSeO$_{3}$.\par

\subsection{Topological phase switching and remnant skyrmion volatility} Next we show $E$ field-driven skyrmion-conical phase switching after MEFC to $T$=54~K and 54.6~K. This $T$ range lies outside the equilibrium SkL phase for $E$=0, yet inside the equilibrium SkL phase for $E$=+5.0~V/$\mu$m - see Fig.~\ref{fig:Efield_sweeping}(a). Therefore, an $E$ field-driven phase switching hallmarked by \emph{in-situ} skyrmion creation and annihilation is expected at these $T$s. Figs.~\ref{fig:Efield_sweeping}(b) and (c) show that a full phase switching is indeed achieved at fixed $T$=54.6~K and $\mu_{0}H$=40~mT. After the initial MEFC to this $T$, sweeping the $E$ field negative suppresses the SkL. The associated SANS intensity falls sharply when the $E$ field falls below +1.0~V/$\mu$m, a remnant metastable SkL exists at $E$=0, with the SkL destroyed completely by -2.5~V/$\mu$m. Concomitantly, the intensity from the coexisting conical phase grows as the SkL is destroyed, with this phase eventually occupying the entire sample by -2.5~V/$\mu$m. When subsequently sweeping the $E$ field back to +5.0~V/$\mu$m, the SkL (conical) phase intensities increase (decrease) for $E>$1.5~V/$\mu$m, and by +5.0~V/$\mu$m have reached their original levels after MEFC. The SANS patterns obtained during the $E$ sweep that evidence the phase switching are shown in Figs.~\ref{fig:Efield_sweeping}(d)-(h) for the skyrmion phase and Figs.~\ref{fig:Efield_sweeping}(i)-(m) for the conical phase.\par

The SANS data presented in Fig.~\ref{fig:Efield_sweeping} provide the hitherto missing microscopic evidence that the volume fractions of the competing skyrmion and conical phases are interchanged directly by sweeping the $E$ field. The reversible phase switching observed at 54.6~K has an associated hysteresis of $\Delta E$=2.5(5)~V/$\mu$m, which increases markedly to 4.0(5)~V/$\mu$m already at the slightly lower $T$ of 54~K [Fig.~\ref{fig:Efield_sweeping}(b) and (c)]. At 50~K, where no equilibrium skyrmion phase can be stabilized in our $E$ field range, the metastable SkL SANS intensity lost on the $E$ field decreasing sweep is not obviously recovered by a subsequent $E$ field increasing sweep.\par

\begin{figure}[t]
\includegraphics[width=0.48\textwidth]{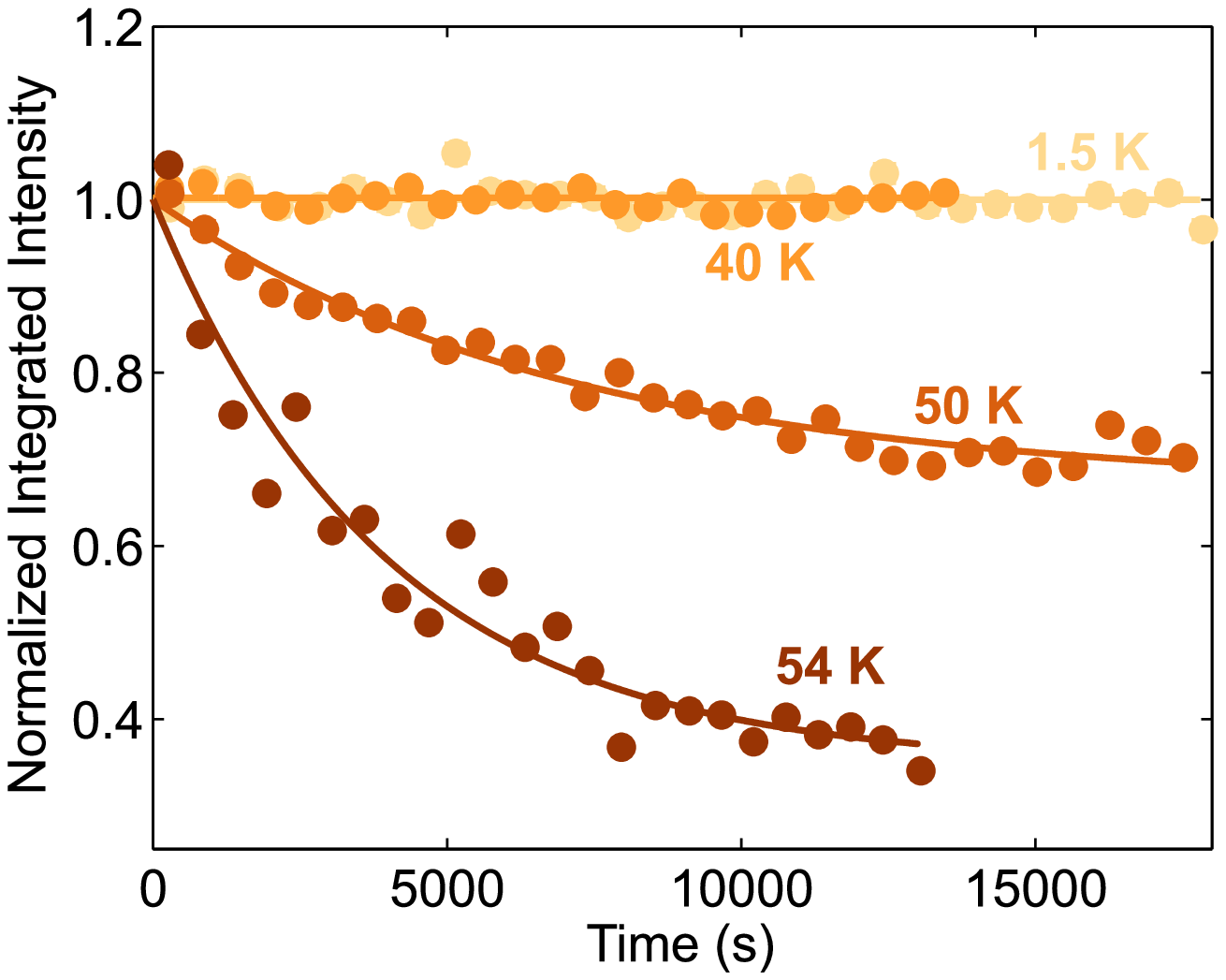}
\caption{Time-dependence of the normalized SANS intensity from the remanent ($E$=0) metastable SkL for $\mu_{0}H=$~40~mT and various $T$s. Each $T$ curve is normalized by the absolute intensity at $t$=0 for easier comparison. Solid lines are fits of each dataset to a simple exponential function.}
\label{fig:volatility}
\end{figure}

Finally we characterized the time ($t$) dependence of the metastable skyrmion state at remnance ($E$=0) in order to learn its stability against thermal agitation. The remnant state was prepared by MEFC to a target $T$, and removing only the biasing $E$ field. Subsequently, the metastable SkL SANS intensity was tracked with $t$. Fig.~\ref{fig:volatility} shows $t$-dependent data obtained at various $T$s, where $t$=0 corresponds to when the $E$ field ramp $\rightarrow 0$ had finished. At 54~K and 50~K, the remnant metastable SkL is thermally volatile, decaying into the conical phase with exponential lifetimes of $\tau$(54~K)=3900(700)~s and $\tau$(50~K)=7000(990)~s, respectively. Thus at high $T$s where the $E$ field driven skyrmion-conical phase switching is achieved, the two states form an asymmetric bistability hallmarked by a thermally-triggered decay of the remnant metastable SkL. It thus follows that any phase change memory functionality considered for this $T$ regime becomes impractical if the required metastable SkL hold-time significantly exceeds $\tau$. In contrast, at the lower $T$s of 40~K and 1.5~K, no discernible decay of the metastable SkL is observed on our experimental time-scale showing $\tau$ to enhance significantly when thermal fluctuations are suppressed.\par

According to existing theories, the microscopic mechanism by which a SkL unwinds into the conical phase is mediated by the nucleation of pinched-off Bloch-point singularities that propagate through the system, and exit at the sample surfaces.~\cite{Mil13,Sch14} If such a mechanism applies to describe the observed decay of metastable skyrmions shown in Fig.~\ref{fig:volatility}, this process is thermally-activated down to at least 50~K, and inhibited for $T$s of 40~K and lower, presumably due to the defect-trapping of Bloch-points. At high $T$s, the timescales of both thermally activated decay (hours), and our SANS measurements (minutes), are longer than those of the observed $E$ field-driven skyrmion destruction and creation processes. Indeed, a recent cryo-Lorentz transmission electron microscopy (cryo-LTEM) study on thin plate Cu$_{2}$OSeO$_{3}$, reports analogous $E$ field-driven modifications of the observed skyrmion density to occur on a timescale of seconds.~\cite{Hua17} Taken together, the data show that skyrmion creation/destruction processes driven by $E$ field can take precedence over thermally-activated processes, implying that it is the $E$ field itself that drives both the nucleation and mobility of the Bloch points required to either destroy or create skyrmion lines. While mechanisms for $E$ field-driven skyrmion creation/destruction have been explored partly in theory~\cite{Wat16,Moc16} a more accurate description of the processes we observe requires theory to further take into account effects due to both thermal fluctuations,~\cite{Kru17} and other factors such as the density of Bloch-point pinning centers. On the experimental side, more detailed $T$-dependent measurements characterizing the crossover between thermally-driven and inhibited decay of the metastable skyrmion state are called for, which can be studied by both reciprocal space techniques such as SANS, and time-resolved real-space imaging by means of cryo-LTEM, or magnetic force microscopy.~\citep{Mil16}\par

At lower $T$ where the metastable skyrmions are more robust, according to our data the near dissipation-free $E$ field-driven switching between two longer-lived bistable states is promised, though significantly larger $E$ fields than used here seem necessary. Different approaches can be considered to remedy this issue. For example, access to much larger $E$ fields can be envisaged by preparing thinner or suitably micro-structured samples. Remaining in the bulk regime, the pinning landscape for Bloch-points may be tuned by partial chemical substitution~\cite{Wu15} possibly providing a route for enhancing the metastable state lifetime at higher $T$s, although $\Delta E$ may become prodigious. The present study may therefore motivate further work aimed at both achieving a truly non-volatile $E$ field-driven topological phase switching, and more deeply understanding the role of $E$ field on the energetics of metastable skyrmion decay and topological protection, which up to now have been studied only as functions of $T$, $\mu_{0}H$ and $t$.~\cite{Mun10,Mil13,Bau16,Wil17,Oik16,Nak17,Kag17a,Kag17,Kar16,Mor17,Kar17,Ban17,Oka17}\par

\section{Summary}
In the magnetoelectric (ME) chiral magnet Cu$_{2}$OSeO$_{3}$ with $T_{c}$=58~K, we have demonstrated the creation of metastable skyrmion states under magnetoelectric field cooling (MEFC) conditions, and the \emph{in-situ} electric ($E$) field-driven phase switching between topologically distinct skyrmion and conical phases. In the high temperature ($T$) regime where phase switching is achieved (close to $T_{c}$), the skyrmion and conical phases form an asymmetric bistability that is characterized by an exponential decay of the remnant metastable skyrmion state on an hour timescale. At lower $T$s the remnant metastable skyrmion lifetime is demonstrably longer-lived, and a phase switching between two truly long-lived bistable states can generally be expected to be driven by sufficiently large $E$ fields. This observed interplay between ME coupling, thermal fluctuations and pinning effects will be directly relevant for anticipated insulating skyrmion phases at room temperature. Therefore these results further furnish the burgeoning perspective for skyrmion-based applications.

\emph{Acknowledgements} Financial support from the Swiss National Science Foundation (SNSF) via the Sinergia network 'NanoSkyrmionics' (grant CRSII5-171003), the SNSF project grants No. 153451, No. 169699, No. 166298, and P2ELP2\textunderscore175278, and the European Research Council grant CONQUEST are gratefully acknowledged. The neutron scattering experiments were performed at the Swiss Spallation Neutron Source (SINQ), Paul Scherrer Institut, Switzerland.

\bibliography{cu2oseo3}

\end{document}